\documentclass[eqsecnum,prd,showpacs,amsmath,aps,nofootinbib,twocolumn]
  {revtex4}  
\usepackage{epsfig}
\usepackage{amssymb}
\usepackage[dvips]{color}
\usepackage[latin1]{inputenc}
\usepackage{graphicx}

\def\ud{\mbox{d}}

\def\beq{\begin{equation}}
\def\eeq{\end{equation}}
\def\beqq{\begin{eqnarray}}
\def\eeqq{\end{eqnarray}}
\def\D{\partial}
\def\beq{\begin{equation}}
\def\eeq{\end{equation}}
\newcommand{\bea}{\begin{eqnarray}}
\newcommand{\eea}{\end{eqnarray}}
\newcommand{\lb}{\label}
\newcommand{\bdm}{\begin{displaymath}}
\newcommand{\edm}{\end{displaymath}}

\newcommand{\I}{{\rm i}}

\makeatletter
\renewcommand*{\@fnsymbol}[1]{\ensuremath{\ifcase#1\or *\or \dagger\or
    \ddagger\or 
   \mathsection\or **\or \dagger\dagger
   \or \ddagger\ddagger \else\@ctrerr\fi}}
\makeatother

\begin{document}
\title{On the Modification of the Cosmic Microwave Background
Anisotropy Spectrum from Canonical Quantum Gravity}

\author{Donato Bini}
\email[E-mail: ]{binid@icra.it}
\affiliation{Istituto per le Applicazioni del Calcolo 
``M. Picone,'' CNR, I-00185 Rome, Italy\\
ICRA, University of Rome ``La Sapienza'', 
00185 Rome, Italy\\
INFN, Sezione di Firenze, Polo Scientifico, Via Sansone 1, 
50019 Sesto Fiorentino, Florence, Italy}

\author{Giampiero Esposito}
\email[E-mail: ]{gesposit@na.infn.it}
\affiliation{Istituto Nazionale di Fisica Nucleare, Sezione di
Napoli, 
Complesso Universitario di Monte S. Angelo, 
Via Cintia Edificio 6, 80126 Napoli, Italy}

\author{Claus Kiefer} 
\email[E-mail: ]{kiefer@thp.uni-koeln.de}
\affiliation{Institut f\"{u}r Theoretische Physik, Universit\"{a}t
zu K\"{o}ln, 
Z\"{u}lpicher Stra{\ss}e 77, 50937 K\"{o}ln, Germany}

\author{Manuel Kr\"amer} 
\email[E-mail: ]{mk@thp.uni-koeln.de}
\affiliation{Institut f\"{u}r Theoretische Physik, Universit\"{a}t
zu K\"{o}ln, 
Z\"{u}lpicher Stra{\ss}e 77, 50937 K\"{o}ln, Germany}

\author{Francesco Pessina}
\email[E-mail: ]{frcs.pessina@gmail.com}
\affiliation{Dipartimento di Fisica,
Complesso Universitario di Monte S. Angelo, 
Via Cintia Edificio 6, 80126 Napoli, Italy}

\date{\today}

\begin{abstract}
We evaluate the modifications to the CMB anisotropy spectrum that
result from a semiclassical expansion of the Wheeler--DeWitt
equation. Recently, such an investigation in the case of a real 
scalar field coupled to gravity, has led to the prediction
that the power at large scales is suppressed. We make here a more
general analysis and show that there is an ambiguity in the choice of
solution to the equations describing the quantum gravitational
effects. Whereas one of the two solutions describes a suppression of
power, the other one describes an enhancement. We investigate possible
criteria for an appropriate choice of solution. The absolute value of
the correction term is in both cases of the same order and currently
not observable. We also obtain detailed formulae for arbitrary
values of a complex parameter occurring in the general solution
of the nonlinear equations of the model. We finally discuss the 
modification of the spectral index connected with the power spectrum
and comment on the possibility of a quantum-gravity induced unitarity
violation.  
\end{abstract}

\pacs{04.60.Ds}

\maketitle

\section{Introduction}

It is well known that black body radiation played a role not only in 
the historical development of quantum theory, 
but also in the formation of the global picture of modern fundamental physics, 
since both the observed microwave background radiation (hereafter CMB) 
in the Universe \cite{gamow,penzias,cobe,wmap,Koma11} 
and the radiation from black holes predicted
on theoretical ground by Hawking \cite{hawking}
have an (approximate) black body spectrum. The former, 
discovered by Penzias and Wilson \cite{penzias}, has revolutionized our 
understanding of cosmology, and the investigation of its anisotropies 
has shed new light on the physics of the very early universe thanks to
the findings of the satellite missions COBE \cite{cobe} and 
WMAP \cite{wmap} and other projects \cite{Koma11}.

The other branch of modern fundamental physics that has motivated
our research is the attempt of building a quantum theory 
of gravity \cite{oup,espo}. 
This was seen for a long time as a logical step, independent of
the ability of performing observations: since gravity couples to the
energy--momentum tensor of matter, and matter fields have a quantum
nature, the scheme where the other fields are quantized whereas
gravity remains classical can only have approximate validity. Although
we cannot yet test physics at the Planck scale, we expect it should
involve a quantum version of gravitation, so that both geometry and
matter fields are quantized. This area is where quantum theory and
gravitational physics meet, leading to the unification of 
guiding principles as well as fundamental interactions, and it involves 
the length scale out of which the present universe evolved.

For a long time, it was thought that quantum gravitational effects,
even when computable in a very accurate way, can hardly be checked
against observations. Over the last two decades, however, 
attempts were made to establish a phenomenology for quantum gravity
(see e.g. \cite{ALMM05}). One particular approach focuses on
quantum gravitational corrections
to the functional Schr\"{o}dinger equation
\cite{singh,BK98,KLM05,kiefer,KK12,calcagni}, 
as they are found from the Wheeler--DeWitt equation 
of canonical quantum gravity \cite{oup}. This will also be the subject
of this paper. An alternative canonical version is loop quantum
gravity. In loop quantum cosmology, 
 analytic formulae for the power spectra of scalar
and tensor perturbations suitable for comparison with observations were obtained
\cite{BCT,calcagni}; alternatively, one can explore the pre-inflationary 
dynamics \cite{agullo}. 
 
Our paper is organized as follows. A brief review of fluctuations
in quantum cosmology is performed in Sec. II, arriving at a coupled
set of nonlinear differential equations. Section III presents the
solution of this
system with and without a mass term, while quantum gravitational
corrections are considered in Sec. IV. Section V studies in detail
the possible predictions of enhancement or suppression of power at
large scales, and Sec. VI is devoted to the spectral index. Concluding
remarks and open problems are discussed in Sec. VII, while the 
Appendix studies in detail the issue of possible violations of unitarity
and how to get rid of them. We use units with 
$\hbar=c=1$ and a redefined Planck mass $m_{\rm P}=\sqrt{3\pi/2G}$. 

\section{Fluctuations in quantum cosmology}

The treatment of small quantum fluctuations on a quantum
Friedmann--Lemaitre--Robertson--Walker background 
was presented by Halliwell and Hawking
\cite{Hall85}. These authors derived effective Schr\"odinger
equations for the various modes, which can be treated independently as
long as the fluctuations remain small.
In canonical quantum gravity, this corresponds to
a Born--Oppenheimer type of approximation \cite{oup}. 
The time in this Schr\"odinger equation is a 
``Jeffreys--Wentzel--Kramers--Brillouin (JWKB) time''
that is defined from the variables of the Friedmann background
(scale factor $a$ and homogeneous field $\phi$).

Using the general Born--Oppenheimer approach presented in Ref.
\cite{singh}, the authors of Ref. \cite{kiefer} have applied this
scheme to fluctuations in quantum cosmology and extended it to the
next order in $m_{\rm P}^{-2}$ to cover the first quantum
gravitational correction terms. It was applied to the anisotropy
spectrum of the CMB in order to derive the modification caused by
these terms.

Let us summarize here the main features of the formalism for the
derivation of the Schr\"odinger equation. For more details, the reader
is referred to \cite{kiefer} and the references therein. We choose a
massive scalar field $\phi$
coupled to gravity in a spatially flat 
Friedmann--Lemaitre--Robertson--Walker universe; its fluctuations
are expanded into Fourier modes with wave vector ${\mathbf k}$
according to
\begin{equation}
\label{eq:1.1}
\delta \phi({\mathbf x},t)=\sum_{k}f_{k}(t)
{\rm e}^{{\rm i}{\mathbf k} \cdot {\mathbf x}}.
\end{equation}
(The scalar-field modes $f_k$ are often called $\delta\phi_k$.)
Strictly speaking, this equation should be replaced by an integral
representation, but we assume here that quantization is performed in a
large box, so that we can take the wave numbers to be discrete for
simplicity. One should, however, bear in mind that the discrete sums
in this paper have to be replaced by integrals if the universe is
infinite. The relation between the discrete and the continuous case is
discussed, for example, in the Appendix of Ref. \cite{PT09}.
 
The full Wheeler--DeWitt equation reads as \cite{Hall85}
\[
\left[\mathcal{H}_0+\sum_{k=1}^{\infty}
\mathcal{H}_{k}\right]
\!\Psi\big(\alpha,\phi,\{f_k\}_{_{k=1}}^{^{\infty}}\big)=0\,, 
\]
where the Hamiltonians $\mathcal{H}_{k}$ of the fluctuation 
modes are given by
\[
\mathcal{H}_k =
\frac{1}{2}\,\text{e}^{-3\alpha}\left[-\,\frac{\partial^2}{\partial
f_k^2} + \Bigl(k^2\,\text{e}^{4\alpha} +
m^2\,\text{e}^{6\alpha}\Bigr)f_k^2\right] .
\]
For the solution, one makes the ansatz
\begin{equation}
\label{eq:1.2}
\Psi \Bigr(\alpha,\phi, \{ f_{k} \}_{k=1}^{\infty}\Bigr)
=\psi_{0}(\alpha,\phi)\prod_{k=1}^{\infty}
{\widetilde \psi}_{k}(\alpha,\phi,f_{k}),
\end{equation}
and performs the redefinition
\begin{equation}
\label{eq:1.3}
\psi_{k}(\alpha,\phi,f_{k}) \equiv \psi_{0}(\alpha,\phi)
{\widetilde \psi}_{k}(\alpha,\phi,f_{k}).
\end{equation}
On writing
\begin{equation}
\label{eq:1.4}
\psi_{k}(\alpha,f_{k})={\rm e}^{{\rm i}S(\alpha,f_{k})},
\end{equation}
and performing the expansion
\begin{equation}
\label{eq:1.5}
S(\alpha,f_{k})=m_{\rm P}^{2}S_{0}+S_{1}+m_{\rm P}^{-2}S_{2}+ \ldots,
\end{equation}
one can derive equations at consecutive orders in $m_{\rm P}$.
At zeroth order in the Planck mass, one writes the
$k$-th component $\psi_{k}^{(0)}$ of the wave function as the product
of a prefactor $\gamma(\alpha)$ with an exponential containing the
phase $S_{1}(\alpha,f_{k})$ according to
\begin{equation}
\label{eq:1.6}
\psi_{k}^{(0)}(\alpha,f_{k})=\gamma(\alpha)
{\rm e}^{{\rm i}S_{1}(\alpha,f_{k})}.
\end{equation}
The JWKB time parameter $t$ is then defined by
\begin{equation}
\label{eq:1.7}
{\partial \over \partial t}=-{\rm e}^{-3 \alpha}
{\partial S_{0}\over \partial \alpha}
{\partial \over \partial \alpha},
\end{equation}
and each $\psi_{k}^{(0)}$ is found to obey a Schr\"{o}dinger equation
of the form
\begin{equation}
\label{eq:1.8}
{\rm i}{\partial \over \partial t}\psi_{k}^{(0)}=
{\cal H}_{k}\psi_{k}^{(0)}.
\end{equation}
For pure exponential inflation, one has $t\equiv \alpha H$.

>From a Gaussian ansatz 
\begin{equation}
\label{eq:1.9}
\psi_{k}^{(0)}(t,f_{k})={\cal N}_{k}^{(0)}(t)
{\rm e}^{-{1\over 2}\Omega_{k}^{(0)}(t)f_{k}^{2}},
\end{equation}
one then finds for ${\cal N}_{k}^{(0)}$ and $\Omega_{k}^{(0)}$ 
a coupled set of nonlinear differential equations,
\begin{eqnarray}
\label{eq:1.10}
{{\rm d}\over {\rm d}t}{\cal N}_{k}^{(0)}(t)&=&-{{\rm i}\over 2}
{\rm e}^{-3\alpha}{\cal N}_{k}^{(0)}(t)\Omega_{k}^{(0)}(t),\\
\label{eq:1.11}
{{\rm d}\over {\rm d}t}\Omega_{k}^{(0)}(t)&=&{\rm i}{\rm e}^{-3\alpha}
\biggr[-\Bigr(\Omega_{k}^{(0)}(t)\Bigr)^{2}\nonumber\\
&& +k^{2}{\rm e}^{4Ht}
+m^{2}{\rm e}^{6Ht}\biggr].
\end{eqnarray}
The second equation has the form of a Riccati equation, which has wide
applications in physics \cite{schuch}. 

The time parameter \eqref{eq:1.7} is the standard Friedmann time $t$
appearing in the standard form of the Robertson--Walker line
element. In spite of this particular choice, the whole formalism is
covariant with respect to time reparametrizations. Using the general
Hamiltonian formalism for canonical gravity, one has in fact
$N^{-1}\partial/\partial t$ appearing in \eqref{eq:1.7} instead of
$\partial/\partial t$, where $N$ is the lapse function (see, e.g. the
detailed explanation of this fact in Sec.~5.4.2 of
\cite{oup}). Consequently, we can use any time parameter we like
(e.g. conformal time instead of Friedmann time) and directly rewrite
all the results in our paper in terms of the new time.

\section{Solution of the nonlinear system 
             with and without mass term} 

In this section, we solve the system of equations \eqref{eq:1.10} and
\eqref{eq:1.11} with and without mass term. We shall here be more
general than in Ref. \cite{kiefer}.
We introduce the variable 
\begin{equation}
\label{eq:2.1}
\xi(t):={k\over H a(t)}.
\end{equation}
The form of (\ref{eq:1.11}) 
suggests defining the dimensionless quantity $\mu:= m/H$.
It is then possible to obtain the general solution in terms of
one unknown parameter, here denoted by $U_{1}$, and the Bessel functions
$J_{\nu}$ and $Y_{\nu}$ with order
\begin{equation}
\label{eq:2.2}
\nu := {1\over 2}\sqrt{9-4 \mu^{2}}.
\end{equation}
This solution is similar in form to the solution of 
the Klein--Gordon equation for a massive scalar field in
de~Sitter space (see e.g. Ref. \cite{GLH89}, or Eq. (51) of Ref. 
\cite{bartolo}, or Sect. 8.3.2 in Ref. \cite{PU}).
The reason is the general connection between the solution of the
classical field and the solution for the function appearing in the
exponent of the Gaussian wave function.

In scenarios of inflation, one assumes that $m<H$ in order to get
fluctuations at super-Hubble scales with quasi-constant amplitude
$H/2\pi$. This yields a real value for $\nu$.

In explicit form, the solution of \eqref{eq:1.11} reads
\begin{eqnarray}
\label{eq:2.3}
& & \Omega_{k}^{(0)}(\xi)= {k^{3}\over H^{2}}
{1\over \xi^{2}(U_{1}Y_{\nu}(\xi)+J_{\nu}(\xi))} 
\nonumber \\
& \times & \Bigr[-{\rm i}U_{1}Y_{\nu+1}(\xi)+{{\rm i}\over 2\xi}
((3U_{1}+2U_{1}\nu)Y_{\nu}(\xi)
\nonumber \\ 
& & \; -2 \xi J_{\nu+1}(\xi)
+(3+2 \nu)J_{\nu}(\xi))\Bigr].
\end{eqnarray}
For the massless case $\mu=0$, (\ref{eq:1.11})
has the general solution  
\begin{eqnarray}
\label{eq:2.4}
\Omega_{k}^{(0)}(\xi)&=&{k^{3}\over H^{2}}
{(U_{1}\cos \xi -\sin \xi){\rm i} \over \xi
[(U_{1}+\xi)\cos \xi+(U_{1}\xi-1)\sin \xi]}
\nonumber \\
&=& {k^{3}\over H^{2}}{{\rm i}\left(J_{{1\over 2}}
-U_{1}J_{-{1\over 2}}\right)\over \xi 
\left[(1-\xi U_{1})J_{{1\over 2}}-(U_{1}+\xi)
J_{-{1\over 2}}\right]}.
\end{eqnarray}
This coincides with the massless limit
of the solution (\ref{eq:2.3}), because then $\nu=3/2$ and standard 
formulae for the Bessel functions of half-odd order show 
the agreement of (\ref{eq:2.3}) with (\ref{eq:2.4}). 

In the following, we shall restrict attention to the massless
case. The massive case can also be dealt with along the following
lines, but is technically much more involved. In Ref. \cite{kiefer}, the
boundary condition was chosen that the quantum state \eqref{eq:1.9}
approaches the free Minkowski vacuum for large wave numbers
$k\to\infty$. This means that for $\xi\to\infty$ one demands that
$\Omega_{k}^{(0)}(\xi)\approx k^3/H^2\xi^2=ka^2$. In the massive case,
this state is known as the Bunch--Davies vacuum \cite{bunchdavies}. 
For this boundary condition, (\ref{eq:2.4}) reduces to the 
solution presented in Eq. (9) of Ref. 
\cite{kiefer}. Formally, it is achieved by setting $U_{1}$ equal to
$-{\rm i}$, and one has then
\begin{equation}
\label{eq:2.8}
\Omega_{k}^{(0)}={k^{3}\over H^{2}}{1\over \xi (\xi-{\rm i})}.
\end{equation}

The general solution
is certainly richer than this one, because it involves ratios of combinations
of Bessel functions of $\xi$, rather than just ratios of
polynomials in the $\xi$ variable.
On writing $U_{1}=\zeta {\rm e}^{{\rm i}\beta}$ for the general case,
we can re-express 
the $\Omega_{k}^{(0)}(\xi)$ in (\ref{eq:2.4}) in the form
\begin{equation}
\label{eq:2.9}
\Omega_{k}^{(0)}(\xi)={k^{3}\over H^{2}}{{\rm i}\over \xi}
{AB^{*}\over |B|^{2}},
\end{equation}
where
\begin{equation}
\label{eq:2.10}
A=\rho + {\rm i} \sigma, \; B=\gamma + {\rm i}\delta,
\end{equation}
and we have defined
\begin{eqnarray}
\label{eq:2.11}
\rho &:= & \zeta \Bigr(\cos (\beta+\xi) +\cos (\beta-\xi)\Bigr)
-2 \sin \xi
\nonumber \\
&=&2 (\zeta \cos \beta \cos \xi- \sin \xi), 
\nonumber\\
\sigma &:=  & \zeta (\sin (\beta+\xi)+\sin (\beta-\xi))
\nonumber\\
&=&2\zeta \sin \beta \cos\xi,
\nonumber\\
\gamma & := & \zeta \Bigr[(\cos (\beta+\xi)+\cos (\beta-\xi))
\nonumber\\
& & \;+\xi (\sin(\beta+\xi)-\sin(\beta-\xi))\Bigr] 
\nonumber \\
& & \; +2 (\xi \cos \xi-\sin \xi) 
\nonumber\\
&=& 2\zeta \left[\cos \beta (\cos \xi+\xi \sin \xi)
-(\sin \xi-\xi \cos \xi) \right],
\nonumber \\
\delta &:=& \zeta \Bigr[(\sin (\beta+\xi)+\sin(\beta-\xi))
\nonumber\\ 
& & \; -\xi (\cos (\beta+\xi)-\cos (\beta-\xi))\Bigr]
\nonumber\\
&=&2\zeta \sin \beta [\cos \xi +\sin \xi].
\end{eqnarray}
As in Ref. \cite{kiefer}, we address here the classical
quantity $\sigma_{k}(t)$ that is related to the quantum mechanical variable
$f_{k}(t)$ of (\ref{eq:1.1}) by the following expectation value taken in a 
Gaussian state $| \psi_{k} \rangle$, i.e.
\begin{equation}
\label{eq:2.15}
\sigma_{k}^{2}(t)= \langle \psi_{k} | f_{k}^{2} | \psi_{k} \rangle
={1\over 2 {\rm Re} \Omega_{k}(t)}.
\end{equation}
In particular, one then finds at the level of approximation connected
with $\psi_k^{(0)}$,
\begin{equation}
\label{eq:2.16}
\sigma_{k}^{(0)}(t)={1\over \sqrt{2}}
\Bigr({\rm Re}\Omega_{k}^{(0)}(t)\Bigr)^{-{1\over 2}}.
\end{equation}
 From the transformation rule
\begin{equation}
\label{eq:2.17}
{{\rm d}\over {\rm d}t}=-\xi H {{\rm d}\over {\rm d}\xi}
\end{equation}
one then gets
\begin{equation}
\label{eq:2.18}
\left | {\dot \sigma}_{k}^{(0)}(t) \right |
= \left | {H \xi \over \sqrt{2}}{{\rm d}\over {\rm d}\xi}
\Bigr({\rm Re}\Omega_{k}^{(0)}(\xi)\Bigr)^{-{1\over 2}} 
\right | .
\end{equation}
\begin{figure}
\includegraphics[scale=0.35]{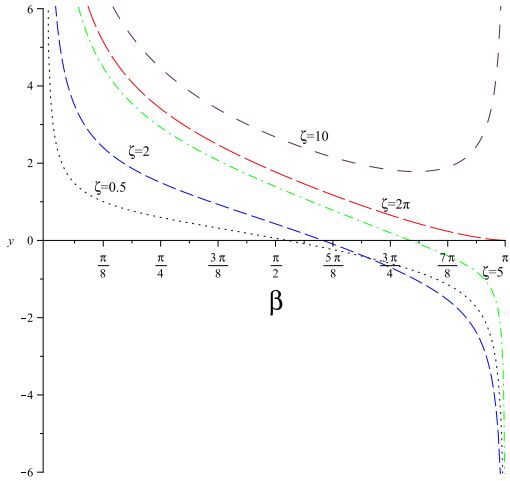}
\caption{The various curves correspond to values of 
$\zeta=0.5,2,5,2\pi,10$ and are ordered so that on the left part of the
figure (say $\beta$ close to $0$) $\zeta$ increases. In other words,
the lower curve corresponds to $\zeta=0.5$, whereas the upper curve
to $\zeta=10$.}
\end{figure}
In our case, this yields
\begin{eqnarray}
\label{eq:2.19}
& & \left | {\dot \sigma}_{k}^{(0)}(t) \right |_{t_{\rm exit}} 
\nonumber\\
&=& {H^{2}\over 2 \sqrt{2}k^{3\over 2}} \left | \xi
\left({(\rho \delta - \gamma \sigma)\over \xi 
(\gamma^{2}+\delta^{2})}\right)^{-{3\over 2}}
{{\rm d}\over {\rm d}\xi}
\left({(\rho \delta - \gamma \sigma)\over \xi 
(\gamma^{2}+\delta^{2})}\right)
\right |_{\xi=2\pi} 
\nonumber \\ & & 
= {2 \sqrt{2}\pi^{2}H^{2}\over k^{3\over 2}}
\left | {\sqrt{\zeta}(\zeta + 2 \pi \cos \beta)
\over \sqrt{\sin \beta} \sqrt{\zeta^{2}+4\pi \cos \beta
+4 \pi^{2}}} \right |.
\end{eqnarray}
The last step follows
because $\xi(t_{\rm exit})=2\pi$ at Hubble-scale crossing 
\cite{kiefer}. Figure $1$ shows a plot of the function whose absolute
value is taken here:
\begin{equation}
f(\zeta,\beta):={\sqrt{\zeta}(\zeta + 2 \pi \cos \beta)
\over \sqrt{\sin \beta} \sqrt{\zeta^{2}+4\pi \cos \beta
+4 \pi^{2}}},
\end{equation}
when $\zeta \in (0,10), \; \beta \in (0,\pi)$. This function
exhibits no relative minima or maxima, and hence there are no
`preferred' values of $\zeta$ and $\beta$ from a mathematical
point of view. The result (\ref{eq:2.19}) can be used to obtain the
power spectrum \cite{kiefer,calcagni} 
\begin{equation}
\mathcal{P}^{(0)}(k):= \frac{k^{3}}{2\pi^2}\left|\delta_{k}(t_{\rm enter})\right|^{2},
\end{equation}
where \cite{pad}
\begin{equation}
\delta_{k}(t_{\rm enter})={4\over 3}
{{\dot \sigma}_{k}(t)\over {\dot \phi}(t)}\Biggr|_{t=t_{\rm exit}},
\end{equation}
such that we get
\begin{equation}
\label{ps-uncorr}
\mathcal{P}^{(0)}(k) =
\frac{64}{9} \pi^{2} \left|f(\zeta,\beta)\right|^{2}
\frac{H^4}{\big|\dot{\phi}(t)\big|^{2}}_{\!\!\!\!t_{\mathrm{exit}}}\!\!\!.  
\end{equation}

\section{Quantum gravitational corrections}

Proceeding with the expansion \eqref{eq:1.5} to the next order, which
contains terms proportional to $m_{\rm P}^{-2}$, we arrive at
a quantum-gravity corrected Schr\"odinger equation of the form
\cite{kiefer} 
\begin{equation}
\label{eq:3.4}
{\rm i}{\partial \over \partial t}\psi_{k}^{(1)}
={\cal H}_{k}\psi_{k}^{(1)}
-{{\rm e}^{3\alpha}\over 2 m_{P}^{2}\psi_{k}^{(0)}}
\left[{({\cal H}_{k})^{2} \over V}\psi_{k}^{(0)}\right]
\psi_{k}^{(1)},
\end{equation}
where 
\begin{eqnarray}
\label{eq:3.2}
V&:=&{\rm e}^{6\alpha}H^{2}, \qquad
W_{k} := k^{2}{\rm e}^{4\alpha}+m^{2}{\rm e}^{6\alpha},\\
{}\label{eq:3.3}
{\cal H}_{k} &:= & {1\over 2}{\rm e}^{-3\alpha}\left[
-{\partial^{2}\over \partial f_{k}^{2}}+W_{k}f_{k}^{2}\right].
\end{eqnarray}
As in Ref. \cite{kiefer}, we make the following Gaussian ansatz for 
the corrected wave functions $\psi_{k}^{(1)}$:
\begin{eqnarray}
\label{eq:3.1}
& & \psi_{k}^{(1)}=\left(N_{k}^{(0)}(t)
+m_{\rm P}^{-2}N_{k}^{(1)}(t)\right)
\nonumber\\ 
& & \; \times
{\rm exp}\left[-{1\over 2}\left(\Omega_{k}^{(0)}(t)
+m_{\rm P}^{-2}\Omega_{k}^{(1)}(t)\right)f_{k}^{2}(t)\right].
\end{eqnarray}
Here, a second term describing a possible violation of unitarity has
been neglected in \eqref{eq:3.4}. Such a term was found in the general
derivation 
\cite{singh} and can be interpreted as follows (see also Ref.
\cite{kief93}). The general Wheeler--DeWitt equation is of the
Klein--Gordon form, not the Schr\"odinger form. Expanding the
conserved Klein--Gordon type current in powers of $m_{\rm P}^{-2}$,
one arrives at order $m_{\rm P}^0$ at the exact conservation of the
Schr\"{o}dinger current, and at order $m_{\rm P}^{-2}$ at a violation of
this conservation by a term that corresponds to the
unitarity-violating term neglected in \eqref{eq:3.1}. 

One can estimate that in most situations the unitarity-violating term
is negligible compared to the correction term in \eqref{eq:3.1}
\cite{singh}. But one can also adopt the following viewpoint. After an
appropriate redefinition of the wave function, unitarity can be
achieved at order $m_{\rm P}^{-2}$ \cite{bertoni}.
Such a procedure was also applied in the context of the
quantum-mechanical Klein--Gordon equation in an external gravitational
field \cite{lammer}. Independently of which argument is used, we shall
no longer consider the unitarity-violating term in the main body of
our paper, but we refer the reader to the Appendix for detailed
calculations aimed at further clarifying this crucial issue.

Inserting the ansatz \eqref{eq:3.1} into \eqref{eq:3.4}, 
one arrives at 
\begin{eqnarray}
\label{eq:3.5}
\; & \; & {\rm i}{{\rm d}\over {\rm d}t}\log \left(
N_{k}^{(0)}+{N_{k}^{(1)}\over m_{P}^{2}}\right)
-{{\rm i}\over 2}\left({\dot \Omega}_{k}^{(0)}
+{{\dot \Omega}_{k}^{(1)}\over m_{P}^{2}}\right)f_{k}^{2}
\nonumber \\
&=& {1\over 2}{\rm e}^{-3\alpha}\biggr \{
\Omega_{k}^{(0)}+{1\over m_{P}^{2}}\left[\Omega_{k}^{(1)}
-{3\over 4V}\left(\Bigr(\Omega_{k}^{(0)}\Bigr)^{2}
-{2\over 3}W_{k}\right)\right] 
\nonumber \\
&+& \left(W_{k}-\left(\Omega_{k}^{(0)}
+{\Omega_{k}^{(1)}\over m_{P}^{2}}\right)^{2}
-{3 \Omega_{k}^{(0)}(W_{k}-(\Omega_{k}^{(0)})^{2})\over
2V m_{P}^{2}}\right)f_{k}^{2}
\nonumber\\
&& +{\rm O}(f_{k}^{4}) \biggr \},
\end{eqnarray}
which can be cast in the form
\begin{equation}
\label{eq:3.6}
\sum_{l=0}^{2}A_{2l}f_{k}^{2l}=0
\end{equation}
with time-dependent coefficients $A_{2l}$. 
Setting, in particular, the overall coefficient $A_{2}$ of 
$f_{k}^{2}$ to zero, one finds the first-order nonlinear 
equation \cite{kiefer}
\begin{eqnarray}
\label{eq:3.7}
& & {\dot \Omega}_{k}^{(1)}(t)=-2{\rm i}{\rm e}^{-3\alpha}
\Omega_{k}^{(0)}(t) \nonumber \\ & & \!\!\! \times
\left(\Omega_{k}^{(1)}(t)-{3\over 4 V(t)}
\left[\Bigr(\Omega_{k}^{(0)}(t)\Bigr)^{2}
-W_{k}(t)\right]\right).
\end{eqnarray} 
Eventually, on defining 
\begin{equation}
\label{eq:3.8}
C := \rho \delta-\gamma \sigma, \quad
D := \rho \gamma+ \sigma \delta,
\end{equation}
this reads as follows (since ${\rm e}^{-\alpha}=H \xi / k$)
\begin{eqnarray}
\label{eq:3.9}
& & {{\rm d}\Omega_{k}^{(1)}\over {\rm d}\xi}
={2{\rm i}\xi \over (\gamma^{2}+\delta^{2})}(C+{\rm i}D)\times 
\nonumber\\
& &  \!\!\!\!\!\!\!\!\!\!\left[
\Omega_{k}^{(1)}+{3\over 4}\left(\mu^{2}+\xi^{2}-\xi^{4}
{(C^{2}-D^{2}+2{\rm i}CD) \over (\gamma^{2}+\delta^{2})^{2}}
\right)\right].
\end{eqnarray}
In Ref. \cite{kiefer}, the desired
solution $\Omega_{k}^{(1)}$ is taken to vanish at late times.
This expresses the idea that quantum gravitational corrections 
should tend to zero at large times, which is certainly in agreement
with observations. In the next section, we shall discuss a subtlety
that arises when solving \eqref{eq:3.9} with this boundary condition. 

\section{Enhancement or suppression of power at large scales?}

Taking as in Ref. \cite{kiefer} the initial state to be the ground state 
of Minkowski spacetime, one has to make the choice $\zeta=1$ and
$\beta={3\over 2}\pi$. For this case, \eqref{eq:3.9} reads
\begin{equation}
\label{eq:3.12}
{{\rm d}\Omega_{k}^{(1)}\over {\rm d}\xi}
={2{\rm i}\xi \over (\xi-{\rm i})}\Omega_{k}^{(1)}
+{3\over 2}\xi^{3}{(2\xi-{\rm i})\over (\xi-{\rm i})^{3}}.
\end{equation}
The corresponding solution of (\ref{eq:3.9}) 
is then inserted into the formula that
generalizes (\ref{eq:2.18}) at the next-to-leading order \cite{kiefer},
\begin{equation}
\label{eq:3.10}
\left | {\dot \sigma}_{k}^{(1)}(t) \right |
=\left | {H \xi \over \sqrt{2}}{{\rm d}\over {\rm d}\xi}
\left[\left({\rm Re}\Omega_{k}^{(0)}+m_{\rm P}^{-2}
{\rm Re}\Omega_{k}^{(1)}(\xi)\right)^{-{1\over 2}}
\right] \right |,
\end{equation}
from which one gets
\begin{equation}
\label{eq:3.11}
\left | {\dot \sigma}_{k}^{(1)}(t) \right |_{t_{\rm exit}}
=|C_{k}| \left | {\dot \sigma}_{k}^{(0)}(t) \right |_{t_{\rm exit}}.
\end{equation}
Eventually, the square of the coefficients $C_{k}$ yields the
corrected power spectrum (see below). 

On requiring $\Omega_{k}^{(1)}(\xi=0)=0$, one finds from \eqref{eq:3.12}
the following exact solution:
\begin{widetext}
\begin{equation}
\label{eq:4.1}
\Omega_{k}^{(1)}(\xi)=-{3\over 8}{\rm e}^{2{\rm i}\xi}
{[1+{\rm Ei}(1,2){\rm e}^{2}]\over (1+{\rm i}\xi)^{2}}
+\Omega_{k}^{(1, {\rm part})}(\xi),
\end{equation}
where
\begin{eqnarray}
\label{eq:4.1bis}
\Omega_{k}^{(1, {\rm part})}(\xi)&=&
{3\over 8}{[-5+6(1+{\rm i}\xi)+4{\rm Ei}
\left(1,2(1+{\rm i}\xi)\right){\rm e}^{2(1+i\xi)}
-4\xi^{2}(1+i\xi)]\over (1+{\rm i}\xi)^{2}}\,,
\end{eqnarray}
\end{widetext}
and ${\rm Ei}(a,z)$ denotes the exponential integral defined by
\begin{equation}
\label{eq:4.2}
{\rm Ei}(a,z) := \int_{1}^{\infty} 
{{\rm e}^{-tz}\over t^{a}} {\rm d}t.
\end{equation}
Here, we have to address the special case
\begin{equation}
\lb{Ei}
{\rm Ei}(1,z)\equiv \Gamma (0,z)\equiv E_1(z),
\end{equation}
cf. Ref. \cite{abra}, Sec. 5.1. From the exact solution 
\eqref{eq:4.1bis}, one finds for the coefficients $C_k$ defined 
in \eqref{eq:3.11} the result
\begin{equation}
\label{eq:5.8}
C_{k}\approx\left(1-{54.37 \over k^{3}}{H^{2}\over m_{P}^{2}}
\right)^{-{3\over 2}}
\left(1+{7.98 \over k^{3}}{H^{2}\over m_{P}^{2}}\right),
\end{equation}
see Figure $2$. They correspond to an enhancement of the 
power spectrum compared to its value when quantum-gravity 
effects are neglected,
\begin{eqnarray}
\label{eq:4.4}
& & \mathcal{P}^{(1)}(k)=\mathcal{P}^{(0)}(k)\,C_{k}^{2}
\nonumber\\
& & \! \sim \mathcal{P}^{(0)}(k)\left[1+{89.54 \over k^{3}}
{H^{2}\over m_{P}^{2}}+{1\over k^{6}}{\rm O}
\left({H^{4}\over m_{P}^{4}}\right)\right]^{2}.
\end{eqnarray}

\begin{figure}
\includegraphics[scale=0.3]{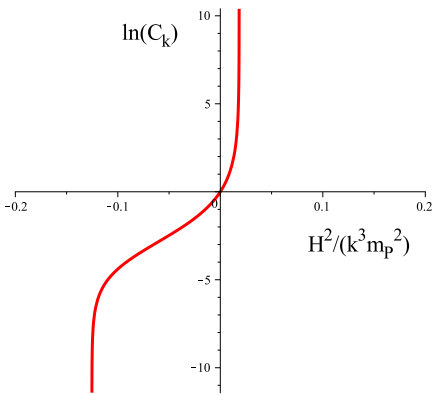}
\caption{The quantity $\ln(C_k)$ is plotted as a function of  
$H^2/(k^3m_{\rm P}^2)$ (see Eq. \eqref{eq:5.8}).}
\end{figure}

In Ref. \cite{kiefer}, another exact solution of \eqref{eq:3.12} is chosen
(although it is there not given in explicit form). The solution has
the same form as in \eqref{eq:4.1bis}, but with the exponential
integral \lb{Ei} replaced by the other exponential integral 
\begin{equation}
\lb{Ei2}
{\rm Ei} (z)= -\int_{-z}^{\infty}\frac{{\rm e}^{-t}}{t}{\rm d}t.
\end{equation}
>From this solution, one finds the value $C_k$ given in Ref. \cite{kiefer},
\begin{equation}
\label{eq:3.13}
C_{k}\approx\left(1-{43.56 \over k^{3}}{H^{2}\over m_{P}^{2}}
\right)^{-{3\over 2}}
\left(1-{189.18 \over k^{3}}{H^{2}\over m_{P}^{2}}\right).
\end{equation}
This solution for $C_{k}$ approaches $1$ at large $k$ 
(i.e. for small scales), but decreases monotonically
to $0$ for small $k$ (i.e. for large scales). In contrast to \eqref{eq:5.8},
this solution thus leads to a suppression of power at large scales,
\begin{eqnarray}
\label{eq:3.14}
& & \mathcal{P}^{(1)}(k)=\mathcal{P}^{(0)}(k)\,C_{k}^{2}
\nonumber \\
& & \!\!\!\!\!\!\!\! = \mathcal{P}^{(0)}(k)
\left[1-{123.83 \over k^{3}}
{H^{2}\over m_{P}^{2}}+{1\over k^{6}}{\rm O}
\left({H^{4}\over m_{P}^{4}}\right)\right]^{2}.
\end{eqnarray}

What is the difference between both solution, that is, between the
different versions of the exponential integral? 
Both solutions ${\rm Ei}(z)$ and $E_1(z)$ assume the value zero if
$\xi=0$. Whereas, however, the solution $E_1(z)$ approaches this value
continuously (see Figure $3$) , 
the function ${\rm Ei}(z)$ makes a jump with size $\pi$
in its imaginary part \cite{abra}, in agreement
with the property \cite{abra}
$$
E_{1}(-x \pm {\rm i}0)=-{\rm Ei}(x) \mp {\rm i} \pi,
$$
where as above $E_{1}(z):=\int_{z}^{\infty}{{\rm e}^{-t}\over t}{\rm d}t$
when $|{\rm arg}z| < \pi$. Imposing continuity as a reasonable
selection criterion for the solution, since our JWKB ansatz for the 
wave function should be differentiable, would entail the choice of
$E_1(z)$ and would thus lead to the prediction of an enhancement of
power at large scales, unlike the prediction of suppression in
Ref. \cite{kiefer}. 

\begin{figure}
\includegraphics[scale=0.35]{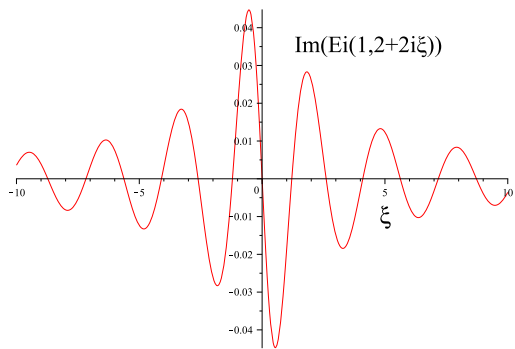}
\caption{Plot of the imaginary part of ${\rm Ei}(1,2i \xi + 2)\equiv
  E_1(2{\rm i}\xi+2$.}
\end{figure}

Let us now revert to Eq. (\ref{eq:3.12}).
Remarkably, by passing to the new variable
\beq
z=1+{\rm i}\xi ,
\eeq
it can be written in the form 
\beq
\label{eq:5.14}
\frac{{\rm d}\Omega_{k}^{(1)}}{{\rm d}z}= 2\left(1-\frac1z  \right)
\Omega_{k}^{(1)} + \frac32 \left(7 -2z -\frac{9}{z} 
+\frac{5}{z^2} -\frac{1}{z^3} \right),
\eeq
and the solution reads as
\begin{eqnarray}
\label{omega1_z}
\Omega_{k}^{(1)}(z)&=&P_1 \frac{{\rm e}^{2z}}{z^2}
+\frac{3}{8z^2}\left[4z^3-8z^2+10z-5\right. 
\nonumber\\
&& \left. +4{\rm e}^{2z} {\rm Ei}(1,2z) \right]\,,
\end{eqnarray}
with $P_1$ determined so that $\Omega_{k}^{(1)}(1)=0$. The passage
to a complex independent variable is therefore very convenient 
in the course of performing and checking our calculations.

Such a solution can be studied graphically by introducing the complex 
polar representation for $z=\rho {\rm e}^{{\rm i}\theta}$ and substituting
it into the definition of $\Omega_{k}^{(1)}(z)$. 
On defining the functions
\begin{equation}
f_{\rho}(\theta):= {\rm Re}\Bigr[\Omega_{k}^{(1)}
(\rho {\rm e}^{{\rm i}\theta})\Bigr], \;
g_{\rho}(\theta):= {\rm Im}\Bigr[\Omega_{k}^{(1)}
(\rho {\rm e}^{{\rm i}\theta})\Bigr], 
\end{equation}
one then finds the behavior displayed in Figures $4$ and $5$. 

\begin{figure}
\includegraphics[scale=0.35]{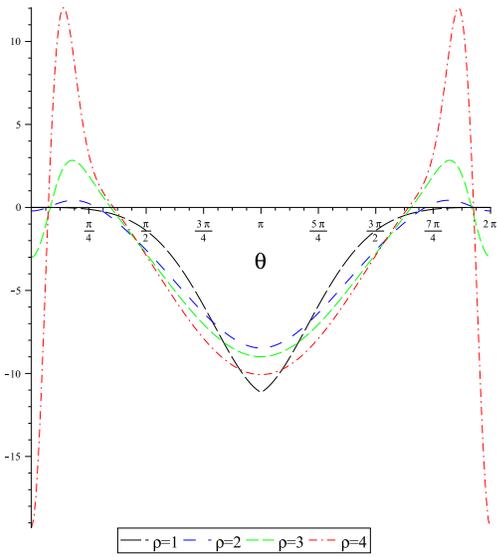}
\caption{Plots of $f_1$, $f_2$, $f_3$, $f_4$ (i.e. fixed values of 
$\rho=1,2,3,4$) as functions of $\theta$. As $\rho$ increases, the curves 
show two enhanced peaks at around $\theta=0$ and $\theta=2\pi$.}
\end{figure}

\begin{figure}
\includegraphics[scale=0.35]{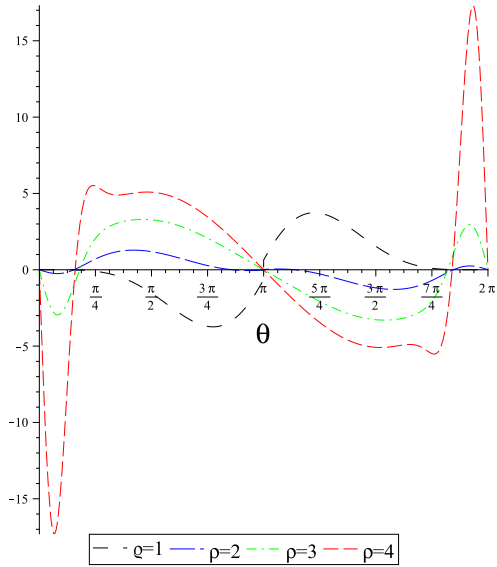}
\caption{Plots of $g_1$, $g_2$, $g_3$, $g_4$ (i.e. fixed values of 
$\rho=1,2,3,4$) as functions of $\theta$. As $\rho$ increases the curves 
show two enhanced peaks at around $\theta=0$ and $\theta=2\pi$.}
\end{figure}

The reader might still be worried by the fact that the limit as
$t \rightarrow \infty$ mentioned before and in Ref. \cite{kiefer}
is only of mathematical interest, because it would correspond to
eternal inflation, which is physically inconsistent. To answer this
question, we begin by noting, from \eqref{eq:2.1}, that 
\begin{equation}
-{1\over H}{{\rm d}\xi \over \xi}={\rm d}t,
\end{equation}
and hence
\begin{equation}
\label{eq:5.17}
\xi(T)={\rm e}^{-H(T-t_{0})}\xi(t_{0}).
\end{equation}
If $T$ denotes the duration of the inflationary stage, it is enough to
impose the boundary condition
\begin{equation}
\Omega_{k}^{(1)}(\xi(T))={\widetilde \varepsilon},
\end{equation}
where, from (\ref{eq:5.17}),
\begin{equation}
\xi(T)=\varepsilon \xi(t_{0}).
\end{equation}
For example, if $\varepsilon$ is taken to be 
$10^{-2},10^{-3}...$ and ${\widetilde \varepsilon}$ is set to $0$,
we still find full agreement with the numerical values 
in Eq. (\ref{eq:5.8}).

If one is instead interested in the deep quantum gravity regime, one 
has to consider values of $T$ so small that $T-t_{0} <0$ 
in (\ref{eq:5.17}). 
It is then appropriate to consider the variable
\begin{equation}
{\widetilde \xi} \equiv {1\over \xi},
\end{equation}
in terms of which Eq. (\ref{eq:5.14}) becomes
\begin{equation}
{{\rm d}\over {\rm d}{\widetilde \xi}}\Omega_{k}^{(1)}
={2{\rm i}\over {\widetilde \xi}^{2}({\rm i}{\widetilde \xi}-1)}
\Omega_{k}^{(1)}+{3\over 2}{({\rm i}{\widetilde \xi}-2)\over
{\widetilde \xi}^{3}(1-{\rm i}{\widetilde \xi})^{3}}.
\label{(5.22)}
\end{equation}
On defining the complex variable $z \equiv 1+{{\rm i}\over \xi}$,
one can then plot $\Omega_{k}^{(1)}$ as a function of such a $z$.
To perform a comparison with Figures $4$ and $5$, we plot in Figures 
$6$ and $7$ the real and imaginary part of the solution of Eq. (\ref{(5.22)})
when the amplitude $\rho$ takes the same values considered in 
Figures $4$ and $5$. The plots corresponding to $\rho=3,4$ are
virtually indistinguishable. The result is not, by itself,
enlightening, but it shows that even the deep quantum gravity 
regime can be further investigated, if necessary. 

\begin{figure}
\includegraphics[scale=0.35]{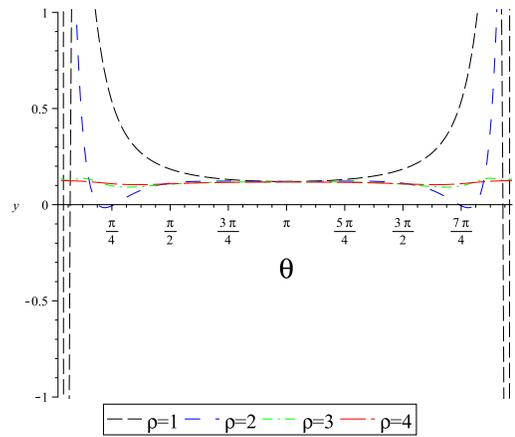}
\caption{Plot of the real part of the solution of (\ref{(5.22)})
when the amplitude $\rho$ of the complex variable 
$z=1+{{\rm i}\over \xi}$ equals $1,2,3,4$.}
\end{figure}

\begin{figure}
\includegraphics[scale=0.35]{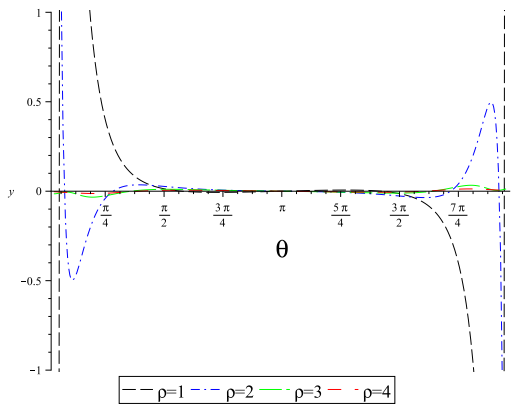}
\caption{Plot of the imaginary part of the solution of (\ref{(5.22)})
when the amplitude $\rho$ of the complex variable 
$z=1+{{\rm i}\over \xi}$ equals $1,2,3,4$.}
\end{figure}

Let us consider finally the general equation (\ref{eq:3.9}). Since the 
dependence on $\zeta$ and $\beta$ complicates matters, we limit our 
considerations to its linearization around $\zeta=1$ and $\beta=3/2\pi$ 
and to the special case $\mu=0$, that is, we look for solutions of the form
\beq
\lb{Omegak1}
\Omega_{k}^{(1)}(\xi)={\widetilde \Omega}_{k}^{(1)}(\xi)
+(\zeta-1)\Omega_{k}^{(1) a}(\xi)+(\beta-3/2\pi)
\Omega_{k}^{(1) b}(\xi)\,,
\eeq
where ${\widetilde \Omega}_{k}^{(1)}$ 
stands for the solution found in \cite{kiefer}.
Substitution of \eqref{Omegak1} into (\ref{eq:3.9}) gives the
following equations:  
\begin{widetext}
\begin{eqnarray}
\frac{{\rm d} \Omega_{k}^{(1) a}}{{\rm d} \xi}&=&
-\frac{2{\rm i}\xi}{({\rm i}-\xi)}\Omega_{k}^{(1) a}
-\frac14 \frac{{\rm i}\xi ({\rm i} \sin \xi 
-2\cos\xi -2{\rm i} \cos^2 \xi\sin \xi 
+2 \cos^3 \xi)}{({\rm i}\sin \xi-3\cos\xi
-4{\rm i} \cos^2 \xi \sin \xi +\cos^3 \xi)(i-\xi)^3}\cdot 
{\mathcal A}(\xi), 
\nonumber\\
\frac{{\rm d} \Omega_{k}^{(1) b}}{{\rm d} \xi}&=&-\frac{2{\rm i}\xi}
{({\rm i}-\xi)}\Omega_{k}^{(1) b} 
+\frac{\xi^2(2\cos^2 \xi -1+2{\rm i} 
\cos\xi \sin \xi)}{4({\rm i}-\xi)^4}\cdot {\mathcal A}(\xi) \,,
\end{eqnarray}
where
\beq
{\mathcal A}(\xi)=12 \xi^4-6\xi^2 +18 {\rm i}\xi
+3+8 {\rm e}^{2{\rm i}\xi}P_{1} +12 {\rm Ei}(1,2(1+{\rm i}\xi))
{\rm e}^{2(1+{\rm i}\xi)},
\eeq
\end{widetext}
and $P_1 = -3/8-3{\rm e}^2{\rm Ei}(1,2)/2$. These equations can be 
studied numerically. Unfortunately, the task turns out to be technically
very hard, but these equations are given here because their 
investigation might shed light on the consequences of choosing another
vacuum for the very early universe. This is the subject of future
investigations.

\section{Observability of the corrections}

We have used the uncorrected Schr\"odinger equation \eqref{eq:1.8} to
arrive at expression \eqref{eq:2.19}, from which we can immediately
obtain the power spectrum \eqref{ps-uncorr}, 
\begin{equation}
\mathcal{P}^{(0)}(k) \propto \frac{H^4}
{\big|\dot{\phi}(t)\big|^{2}}_{\!\!\!\!t_{\mathrm{exit}}}.
\end{equation}
This corresponds -- apart from a dimensionless constant, which is not
relevant for the following discussion -- to the standard power
spectrum of scalar cosmological perturbations \cite{calcagni}, 
\begin{equation}
{\cal P}_{s}^{(0)}(k)={G \over \epsilon \pi}H^{2},
\end{equation}
where we have used the first slow-roll parameter defined as
\begin{equation}
\epsilon=-{{\dot H}\over H^{2}} = \frac{4\pi G \,
\big|\dot{\phi}(t)\big|^{2}_{t_{\mathrm{exit}}}}{H^2}\,.
\end{equation}
As we have seen in \eqref{eq:4.4} and \eqref{eq:3.14}, the
quantum-gravitationally corrected Schr\"odinger equation leads to a
modification of the power spectrum by a correction function $C_k$,
such that we can translate this modification also to the standard
power spectrum in the following way: 
\begin{equation}
{\cal P}_{s}^{(1)}(k)={\cal P}_{s}^{(0)}(k)\,C_{k}^{2}\,.
\end{equation}
Along the lines of \cite{calcagni}, we write
\begin{equation} \label{Ckdelta}
C_{k}^{2}=1+\delta^\pm_{\mathrm{WDW}}(k) + {1\over k^{6}}\,{\mathcal
O}\!\left(\left({H \over m_{\mathrm{P}}}\right)^{4}\right), 
\end{equation}
where $\delta^\pm_{\mathrm{WDW}}(k)$ either takes the form 
\begin{equation} \label{delta+}
\delta^+_{\mathrm{WDW}}(k)={179.09 \over k^{3}}\left({H \over
m_{\mathrm{P}}}\right)^{2}, 
\end{equation}
which follows from \eqref{eq:4.4}, or the form resulting from \eqref{eq:3.14}
\begin{equation} \label{delta-}
\delta^-_{\mathrm{WDW}}(k)=-\,{247.68 \over k^{3}}\left({H \over
m_{\mathrm{P}}}\right)^{2}. 
\end{equation}
The basic equations in the theory of the spectral index $n_s$ and its
running $\alpha_s$ involve the slow-roll parameters
\cite{calcagni} 
\begin{equation}
\eta := -{\ddot \phi \over H {\dot \phi}}\quad\text{and}\quad\Xi^2 :=
{1\over H^{2}}{{\rm d}\over {\rm d}t}{\ddot \phi \over {\dot \phi}}.
\end{equation}
There is thus a change of sign in front of
$\delta^\pm_{\mathrm{WDW}}$ with respect to the discussion in
\cite{calcagni}, that is, 
\begin{equation}
n_{s}-1 := {{\rm d}\log {\cal P}_{s} \over 
{\rm d}\log k} \approx 2\eta -4\epsilon -3 \delta^\pm_{\mathrm{WDW}}
\end{equation}
and
\begin{equation}
\alpha_{s} := {{\rm d}n_{s} \over {\rm d}\log k}
\approx 2 (5 \epsilon \eta-4 \epsilon^{2}-\Xi^{2})
+9 \delta^\pm_{\mathrm{WDW}}\,,
\end{equation}
where use has been made of the approximate 
formula \cite{calcagni}
\begin{equation}
{{\rm d}\over {\rm d}\log k} \approx {1\over H}
{{\rm d}\over {\rm d}t}\,,
\end{equation}
jointly with the equations of motion.

Giving up our definition that $k$ is dimensionless and reinserting a
reference wavenumber, which can either correspond to the largest
observable scale, $k_\mathrm{min} \sim 1.4\times
10^{-4}\,\text{Mpc}^{-1}$ \cite{calcagni}, or to the pivot scale used
in the WMAP9 analysis, $k_0 = 0.002\,\text{Mpc}^{-1}$ 
\cite{calcagni,Hinshaw12}, we now write $k/k_{\mathrm{min}}$ or $k/k_0$,
respectively, instead of $k$. Since the ratio $H/m_{\mathrm{P}}$ has
to be smaller than about $3.5 \times 10^{-6}$ 
because of the observational bound
on the tensor-to-scalar ratio $r < 0.11$ for $k_0 = 0.002\,
\text{Mpc}^{-1}$ from the Planck 2013 results
\cite{baumann, Planck13}, we find that
for $k \rightarrow k/k_0$ the absolute value of the
quantum-gravitational correction is limited by 
\begin{equation}
\left|\delta^+_{\mathrm{WDW}}(k_0)\right| \lesssim 2.2 
\times 10^{-9},\;\; \left|\delta^-_{\mathrm{WDW}}(k_0)\right| 
\lesssim 3.0 \times 10^{-9}, \nonumber
\end{equation}
while with the replacement $k \rightarrow k/k_\mathrm{min}$ 
this limit is further weakened:
\begin{equation}
\left|\delta^+_{\mathrm{WDW}}(k_0)\right| \lesssim 7.5 
\times 10^{-13},\;\; \left|\delta^-_{\mathrm{WDW}}(k_0)\right| 
\lesssim 1.0 \times 10^{-12}. \nonumber
\end{equation}
The difference for $\delta^-_{\mathrm{WDW}}$ compared to
\cite{calcagni} stems from fact that our upper bound for the ratio
$H/m_{\mathrm{P}}$ is weaker than the upper bound derived in
\cite{calcagni} from a different assumption. Furthermore, we have used the
exact values of \eqref{delta+} and \eqref{delta-} instead
of the approximate value $10^3$ as in \cite{calcagni}. 

By comparing the quantum-gravitational corrections to the spectral
index $n_s$ and its running $\alpha_s$ derived above with the values
determined from the WMAP9 data, $n_s = 0.9608 \pm 0.0080$ and $\alpha_s
= -0.023 \pm 0.011$ (using the WMAP9+eCMB+BAO+$H_0$ dataset in both cases) \cite{Hinshaw12},  and the 2013 results of the Planck mission, $n_s = 0.9603 \pm 0.0073$ and $\alpha_s
= -0.013 \pm 0.009$ (using additionally the WMAP polarization data in both cases) \cite{Planck13}, we see that our
corrections are completely drowned out by the statistical uncertainty
in the data. Furthermore, one can already rule out that further improvements of the
statistics of Planck data and future satellite missions to measure the CMB anisotropies more precisely
will push these corrections into observable regions, as the main
source for statistical uncertainty on large scales in the anisotropy
spectrum is cosmic variance, which is given in terms of spherical
multipoles $\ell = 2k/k_\mathrm{min}$ by (see e.g.~\cite{knox}) 
\begin{equation}
\text{Var}_{\mathcal{P}_s}(\ell) = \frac{2}{2\ell+1}\,\mathcal{P}^2_s(\ell)\,.
\end{equation}
Defining $\ell_0$ as the multipole corresponding to the pivot scale
$k_0$ given above, $\ell_0 \approx 29$, the conclusion of the detailed
discussion in \cite{calcagni}, which determines the region in the
$\left(\ell,\mathcal{P}_s(\ell)/\mathcal{P}_s(\ell_0)\right)$ plane
that is affected by cosmic variance can essentially be carried over to
the present discussion. The only difference is the sign change for
$\delta^+_{\mathrm{WDW}}(k)$ and a very slight suppression of the
correction if one does not approximate the pre-factors in 
\eqref{delta+} and \eqref{delta-}. The order of magnitude of the
correction stays the same and therefore also the conclusion that the
quantum-gravitational correction is entirely negligible compared to
the error induced by cosmic variance remains. 

We can repeat, however, for $\delta^+_{\mathrm{WDW}}$ the analysis to
determine an upper bound on 
the energy scale of inflation, which was presented in \cite{kiefer}
for $\delta^-_{\mathrm{WDW}}$. Instead
of assuming that $C_k^2$ has to be greater than about $0.95$ in order
to be compatible with the observation that the anisotropy spectrum
deviates from a scale-invariant spectrum by less than $5\,\%$, we
introduce the upper bound that $C_k^2$ has to be smaller than $1.05$
at $k/k_0 \sim 1$. From equation \eqref{Ckdelta} and \eqref{delta+},
we then immediately obtain the upper bound 
\begin{equation}
H \lesssim 1.67 \times 10^{-2}\,m_{\mathrm{P}} \approx 4.43 
\times 10^{17}\,\text{GeV}.
\end{equation}
For $\delta^-_{\mathrm{WDW}}$ the upper bound is \cite{kiefer}
\begin{equation}
H \lesssim 1.42 \times 10^{-2}\,m_{\mathrm{P}} \approx 3.76 
\times 10^{17}\,\text{GeV}.
\end{equation}
Both constraints are clearly weaker than the already existing
observational limits of about $H \lesssim 10^{15}\,\text{GeV}$, but it
reassures that the present approach is consistent with these limits. 

As a final remark we want to add that the general discussion in
\cite{calcagni} on non-Gaussianities in the squeezed limit arising
from quantum-gravitational corrections is not affected by our results
in the present paper, and we therefore have to conclude that, {\it at the
present state, one cannot expect to see any effect from Wheeler--DeWitt
quantum cosmology in the bispectrum}.  

\section{Concluding remarks}

We have studied the quantum gravitational corrections terms to the CMB
anisotropy spectrum as they are found from a Born--Oppenheimer type of
approximation from the Wheeler--DeWitt equation. We have, in
particular, discussed a subtlety concerning the central equation
\eqref{eq:3.12}. Imposing the boundary condition $\Omega_k^{(1)}\to 0$
for $\xi\to 0$ (corresponding to the absence of quantum-gravity
effects at late times), we have found that there exist two solutions.
One of them, which is related to the exponential integral $E_1(z)$, is
continuous in this limit, whereas the other one, which is related to
the exponential integral ${\rm Ei} (z)$ makes a jump in the imaginary
part of size $\pi$. Both solutions are in accordance with the
requirement that quantum-gravity effects are unobservable at late
times. The numerical corrections to the CMB spectrum are of the same
order in both cases. There is, however, a qualitative
difference. Whereas the continuous solutions leads to an enhancement
of power at large scales, the discontinuous solution leads to a
suppression. If we adopt continuity as a condition for the allowed
solutions, bearing in mind that the JWKB ansatz for the wave
function should be differentiable, 
we have to predict an enhancement of power, unlike 
\cite{kiefer} where a suppression is predicted. 

So far, these corrections to the CMB power spectrum are
too small to be observed. Nevertheless, {\it their size 
is much bigger than corresponding corrections in laboratory
situations}. One can thus express the hope that it might eventually be
possible to test them in a cosmological setting.

At a field-theoretic level, an interesting issue is whether the
various choices of amplitude $\zeta$ and phase $\beta$ 
in Sec. III and V can describe physically 
relevant choices of vacuum other than
the Bunch--Davies vacuum. A further issue seems to be
whether our quantum cosmological calculations can be used to
test the recent theoretical prediction of circles in the CMB
\cite{penrose}. These are subjects for future work.

For a full-fledged investigation, one has to repeat our whole
analysis by using the
gauge-invariant formalism of cosmological perturbation theory
\cite{mukhanov,langlois}. While this will be postponed to a later
paper, we emphasize that the main qualitative features and open
problems are clearly seen already in the more restricted analysis
presented here.

Modifications of the power spectrum at large scales can also emerge
from other situations. 
Inflationary models with 
$\Omega_{\rm m}\neq1$ lead to 
an enhancement of power for low multipoles \cite{Starobinsky},
both for open and closed models. A suppression of power is predicted
from the consideration of a self-interacting scalar field \cite{KS10},
a model of just-enough inflation \cite{RS12}, or the introduction of
non-commutative geometry \cite{TMB03}. 
Yet other work \cite{laura}  
has studied the nonlocal entanglement of the Hubble
volume with modes and domains beyond the horizon, finding that this
induces a dipole and quadrupole contribution in the CMB.  
The great challenge is, of course, to distinguish these contributions
in possible observations.

\acknowledgments
G.\,E. is grateful to the Dipartimento di Fisica of
Federico II University, Naples, for hospitality and support, and 
to Giovanni Venturi for bringing the work in Ref.~\cite{bertoni} to
his attention. C.\,K. thanks Dominik Schwarz for
discussions. M.\,K. acknowledges support from the Bonn--Cologne 
Graduate School of Physics and Astronomy. 

\appendix
\section{The unitarity violation issue}
 
We are here going to see how the unitarity
issue is solved with an alternative 
approach that enables us to explicitly verify that no unitarity 
violations occur. For this purpose, relying upon Ref. \cite{bertoni}, 
we start from the equation (see also Sec.~5.4.1 in \cite{oup})
\beq\label{prima}
\hat{H}\Psi(a,\phi)=\left(\frac{\hbar^2}{2m_{\rm P}^2}
\frac{\partial^2}{\partial a^2}+m_{\rm P}^2V(a)+\hat{H}_M\right)\Psi(a,\phi)=0,
\eeq
where $\hat{H}_M$ denotes a Hamiltonian depending generically 
on matter fields $\phi$ whose mass is light, $m_\phi\ll m_{\rm P}$.
By virtue of this inequality, introducing the Born--Oppenheimer 
(hereafter BO) approximation 
\beq
\Psi(a,\phi)=\psi(a)\chi(a,\phi),
\eeq
we obtain 
\begin{small}
\beq\label{WDW}
\frac{1}{2m_{\rm P}^2}\left(\psi\frac{\partial^2}{\partial a^2}\chi 
+2\frac{\partial \psi}{\partial a}\frac{\partial \chi}{\partial a}
+\chi\frac{\partial^2}{\partial a^2}\psi\right)+\psi V_G\chi 
+ \psi\hat{H}_M\chi =0, 
\eeq
\end{small}
with $V_G=m_{\rm P}^2V(a)$. In our BO approximation or, in other terms,
adiabatic approximation, 
we can neglect the term $\frac{\partial^2}{\partial a^2}\chi$ 
because it varies slowly with respect to the scale factor $a$. 
Hereafter, in order to agree with the notation of
\cite{Brout}, we will  
use the Dirac-like notation so that $\chi (a,\phi )
=\langle a,\phi|\chi\rangle$. Thus, evaluation of $\langle\chi|$ 
on the equation obtained from (\ref{WDW}) upon 
replacing $\chi(a,\phi)$ with $\langle a,\phi|\chi\rangle$ yields
\beq\label{WDW2}
\frac{1}{2m_{\rm P}^2}\frac{\D^2}{\D a^2}\psi + V_G\psi 
+ \langle\hat{H}_M\rangle\psi 
+ \frac{1}{m_{\rm P}^2}\left\langle\chi\left|\frac{\D}{\D a}
\right|\chi\right\rangle\frac{\D\psi}{\D a}=0,
\eeq
where we have assumed $\left\langle\chi|\chi\right\rangle
=\int\ud\phi\chi^*\chi=1$.
Now we obtain $V_G\psi$ from the latter, and substituting it 
into (\ref{WDW}) gives
\beqq\label{Brout}
\left[\hat{H}_M-\langle H_M\rangle\right]|\chi\rangle &
+&\frac{1}{m_{\rm P}^2}\frac{\D\ln\psi}{\D a}
\left[\left.\left|\frac{\D}{\D a}\right|\chi\right\rangle\right.
\nonumber \\
&-&\left.\left\langle\chi\left|\frac{\D}{\D a}\right| \chi  
\right\rangle|\chi\rangle \right]=0.
\eeqq
We can obtain the same result by keeping all terms and doing the 
approximation after the calculation. In fact, from (\ref{WDW}), if
we evaluate $\langle\chi|$ on the left-hand side and add and 
subtract $\psi\times(\left\langle\chi|\frac{\D}{\D a}|
\chi\right\rangle)^2$, we get
\beq\label{Venturi}
\left[\frac{\hbar^2}{2m_{\rm P}^2}D^2+V_G+\langle\hat{H}_M\rangle\right]\psi
=-\frac{\hbar^2}{2m_{\rm P}^2}\langle\bar{D^2}\rangle\psi,
\eeq
where we have introduced the operators
\beq
D:= \frac{\D}{\D a}+\I A, \qquad 
\bar{D}:= \frac{\D}{\D a}-\I A, 
\eeq
\beq
A:=-\I\left\langle\chi\left|\frac{\D}{\D a}\right|\chi\right\rangle.
\eeq
Following \cite{BV}, the condition 
$\left\langle\chi|\frac{\D}{\D a}|
\chi\right\rangle=0$  can be
seen as a `gauge' choice and it implies no restriction by virtue of 
the `gauge invariance' of the full wave function. 
We choose $\left\langle\chi|\frac{\D}{\D a}|\chi\right\rangle=0$
as Eq. (\ref{WDW2}) suggests because the equation we require for the 
semiclassical theory of gravity is that in which classical gravity is
driven by the mean energy of matter. \\
Evaluation of $\langle\chi|$ on Eq. (\ref{Venturi}) and insertion 
into Eq. (\ref{prima}) leads to 
\beq\label{Venturi2}
\psi(\hat{H}_M-\langle\hat{H}_M\rangle)\chi 
+\frac{\hbar^2}{m_{\rm P}^2}D\psi\bar{D}\chi=-\frac{\hbar^2}{2m_{\rm P}^2}
\psi(\bar{D}^2-\langle\bar{D}^2\rangle)\chi.
\eeq
This equation becomes equal to (\ref{Brout}) if we define 
the new wave functions 
\beqq
\tilde{\psi}&:= & \exp\left[\I\int^a A\ud a'\right]\psi, \\
\tilde{\chi}&:= & \exp\left[-\I\int^a A\ud a'\right]\chi, \\
\tilde{\Psi}&:= &\tilde{\psi}\tilde{\chi}=\Psi.  
\eeqq
At this stage, the BO approximation is imposed by neglecting the 
terms $\langle\bar{D}^2\rangle$ and $\langle\bar{D}^2-
  \langle\bar{D}^2\rangle\rangle$ that are related to fluctuations.\\
The next step is to require that $\psi$ should be the JWKB solution 
of Eq. (\ref{WDW2})
\beq
\psi=\frac{1}{N}\exp\left[\frac{\I S}{\hbar}\right],
\eeq
where $S$ is the classical action defined by the trajectories that solve 
Eq. (\ref{WDW2}), and 
\beq
N=\sqrt{\left|\frac{\D S}{\D a}\right|}=\sqrt{m_{\rm P}^2 \dot{a}},
\eeq
as one can obtain from the JWKB method. Now, substituting this relation 
in (\ref{Brout}), we obtain an equation for $|\chi\rangle$
\beq
\left[\hat{H}_M-\langle H_M\rangle\right]|\tilde{\chi}\rangle 
- \frac{\I}{m_{\rm P}^2}\frac{\D S}{\D a}\frac{\D}{\D a}|\tilde{\chi}\rangle=0.
\eeq
The latter might be regarded as a Schr\"{o}dinger equation for the matter 
wave function 
\beq
\left(H_M-\I\hbar\frac{\D}{\D \eta}\right)|\chi_s\rangle=0,
\eeq
where we have introduced the time derivative through
\beq
\I\hbar\frac{\D}{\D \eta}:= -\I\frac{\hbar}{m_{\rm P}^2}
\frac{\D S}{\D a}\frac{\D}{\D a},
\eeq
and we have defined the new wave function 
\beq\label{chi}
|\tilde{\chi}_s\rangle := \exp\left[-\frac{\I}{\hbar}\int^\eta
\langle\hat{H}_M\rangle\ud\eta'\right]|\tilde{\chi}\rangle.
\eeq
So we note that at the zeroth stage of the JWKB approximation one obtains 
the usual evolution equation for matter 
(Schr\"{o}dinger equation or, in the field-theoretic case,
Tomonaga--Schwinger equation) \cite{oup}. 

One may now search for a possible violation of unitary evolution 
for this matter wave function. In order to check this, we perform all 
the approximations after the calculation. 
On considering all terms, (\ref{Brout}) becomes 
\bea\label{chis}
& & \left(H_M-\I\hbar\frac{\D}{\D \eta}\right)|\tilde{\chi}_s\rangle =
\nonumber\\ \!\!\!\!\! & & 
-\frac{\hbar^2}{m_{\rm P}^2}\left[-\frac{\D \log N}{\D a}\bar{D}+\frac{1}{2}
(\bar{D}^2-\langle\bar{D}^2\rangle)\right]|\tilde{\chi}_s\rangle.
\eea
Now we can investigate a possible violation of unitarity from the relation 
\beqq\label{unitaritycalculation}
 &\;&\I\frac{\D}{\D \eta}\langle\tilde{\chi}_s|\tilde{\chi}_s
\rangle =\I\int\ud\phi\left[\langle\tilde{\chi}_s|
\frac{\D}{\D\eta}\tilde{\chi}_s\rangle-C.C.\right]
\nonumber \\ 
&=&\int \ud\phi\left[\langle\chi|\left(\hat{H}_M
-\frac{\hbar^2}{m_{\rm P}^2}\frac{\D \log N}{\D a}\bar{D}\right.\right. 
\nonumber \\
&+& \left.\left.\frac{\hbar^2}{2m_{\rm P}^2}(\bar{D}^2
-\langle\bar{D}^2\rangle\right)|\chi\rangle -C.C.\right]=0,
\eeqq
because we have to neglect $\langle\bar{D}^2\rangle$ and 
$\langle\bar{D}^2-\langle\bar{D}^2\rangle\rangle$ in the BO approximation. 
Thus, unless one considers a nonHermitian Hamiltonian operator, there is 
no violation of unitarity.

Now we want to find a relation between our approach and this one.
It is indeed possible to rewrite the BO approximation in the form 
\beq
\Psi=\exp\left[\frac{\I S}{\hbar}\right],
\eeq
and expanding S in powers of $m_{\rm P}^2$ 
\beqq
S&=&m_{\rm P}^2S_0+S_1+\frac{1}{m_{\rm P}^2}S_2+\mathcal{O}
\left(\frac{1}{m_{\rm P}^4}\right)  \\
  &=&\left[m_{\rm P}^2S_0(a)+\frac{1}{m_{\rm P}^2}\sigma_2(a)\right]
\left[S_1(a,\phi)+\frac{1}{m_{\rm P}^2}\eta_2(a,\phi)\right], \nonumber
\eeqq
where we have decomposed $S_2(a,\phi)=\sigma_2(a)+\eta_2(a,\phi)$.
Thus, the wave functional becomes
\beqq
\Psi &\approx &\left(\frac{1}{N_K(a)}\exp\left[
\frac{\I m_{\rm P}^2}{\hbar}S_0
+\frac{\I}{\hbar m_{\rm P}^2}\sigma_2\right]\right)
\nonumber \\
&\;& \left(N_K(a)\exp\left[\frac{\I S_1}{\hbar}
+\frac{\I\eta_2}{\hbar m_{\rm P}^2}\right]\right)  
\nonumber \\
&=&\tilde{\psi}_K\tilde{\chi}_K.
\eeqq
In order to obtain the desired equations, we have to substitute these 
expressions for $\tilde{\psi}_K$ and $\tilde{\chi}_K$ in 
(\ref{Venturi}) and (\ref{Venturi2}). From (\ref{Venturi}) we have three 
equations proportional to $\mathcal{O}(m_{\rm P}^2)$, 
$\mathcal{O}(m_{\rm P}^0)$, $\mathcal{O}(m_{\rm P}^{-2})$, respectively,
\beqq
&\;&-\frac{1}{2}{S_0'}^2+V=0, \\
&\;&{\langle\hat{H}_M\rangle}_0-\I\hbar\frac{N_K'S_0'}{N_K}
+\I\hbar\frac{S_0''}{2}=0,\\  
&\;&\frac{\hbar^2}{2}\left(2\frac{{N_K'}^2}{N_K^2}
-2\frac{S_0'\sigma_2'}{\hbar^2}-\frac{N_K''}{N_K}\right)
\nonumber \\
&+&{\langle\hat{H}_M\rangle}_{-2}=-\frac{\hbar^2}{2}{
\langle\bar{D}^2\rangle}_0 ,
\eeqq
where a prime denotes a derivative with respect to $a$.
Analogously, from (\ref{Venturi2}) one obtains
\beqq
&\;& H_M^0-\langle\hat{H}_M\rangle_0+\hbar\I S_0'\left(\frac{N_K'}{N_K}
+\frac{\I S_1'}{\hbar}\right)=0, \\
&\;& H_M^{-2}-\langle\hat{H}_M\rangle_{-2}-\hbar^2\frac{N_K'}{N_K}
\left(\frac{N_K'}{N_K}+\frac{\I S_1'}{\hbar}\right) \\
&\;& -S_0'\eta_2' = \frac{\hbar}{2}\left(\langle\bar{D}^2\rangle_0 
-\frac{N_K''}{N_K}-\I\frac{S_1''}{\hbar}+\frac{{S_1''}^2}{\hbar}
-2\frac{\I S_1'}{\hbar}\frac{N_K'}{N_K}\right), 
\nonumber
\eeqq
to $\mathcal{O}(m_{\rm P}^0)$ and $\mathcal{O}(m_{\rm P}^{-2})$ respectively 
(where we have defined the $c$-number $H_M$ by 
$\hat{H}_M|\tilde{\chi}_K\rangle=H_M|\tilde{\chi}_K\rangle$).
By comparing terms $\mathcal{O}(m_{\rm P}^0)$, $\mathcal{O}(m_{\rm
  P}^{-2})$ we have 
\beqq\label{correzione0}
&\;&H_M^0+\frac{\I\hbar S_0''}{2}-S_0'S_1'=0,\\ \label{correzione2}
&\;&-S_0'\eta_2 '+ H_M^{-2}-S_0'\sigma_2 '+\I\hbar\frac{S_1''}{2}
-\frac{{S_1'}^2}{2}=0,
\eeqq
respectively, where we have expanded $H_M=H^2_M+H^0_M+H^{-2}_M
+\ldots$. As one can see,  
\eqref{eq:1.8} and \eqref{eq:3.4} are identical to 
(\ref{correzione0}) and (\ref{correzione2}) if we choose $H_M=H_k$. 
To perform the calculation of the possible violation of unitarity, 
it is extremely useful to note that the condition 
$\langle\chi|\bar{D}|\chi\rangle=0$ becomes
\beqq
\langle\chi|\bar{D}|\chi\rangle &=& \langle\tilde{\chi}|
\frac{\D}{\D a}|\tilde{\chi}\rangle =0,
\nonumber \\
\langle\tilde{\chi}_{Ks}|\frac{\D}{\D a}|\tilde{\chi}_{Ks}\rangle 
&=&\int\ud\phi\left[\tilde{\chi}^*_K\left(\frac{N_K'}{N_K}
+\I\frac{S_1'}{\hbar}+\I\frac{\eta_2'}{m_{\rm P}^2\hbar}\right)
\tilde{\chi}_K\right]
\nonumber \\
&=&0.
\eeqq
In the same way, as in (\ref{chis}), for the $\mathcal{O}(m_{\rm P}^{-2})$ 
we obtain
\beqq
&\;&\left[\left(H_M^0+\frac{1}{m_{\rm P}^2}H_M^{-2}\right)
-\I\frac{\D}{\D\eta}\right]|\tilde{\chi}_{Ks}\rangle 
\\
&=&\left[\frac{\hbar^2}{m_{\rm P}^2}\left(\frac{{N_K'}^2}{N_K^2}
+\I\frac{N_K'S_1'}{\hbar}\right)\right. 
\nonumber \\ 
&+& \frac{\hbar^2}{2m_{\rm P}^2}\biggr (\left\langle\tilde{\chi}\left|
\left(\frac{\D^2}{\D a^2}\right)\right|\tilde{\chi}\right\rangle_0
\nonumber \\
&-& \left. \frac{N_K''}{N_K}-\I\frac{S_1''}{\hbar}
+\frac{{S_1'}^2}{\hbar^2}
-2\I\frac{N_K'S_1'}{\hbar N_K}\right)\biggr]|\tilde{\chi}_{Ks}\rangle.
\nonumber
\eeqq
Performing the same calculation as in (\ref{unitaritycalculation}), 
we finally obtain the same result
\beqq
&\;&\I\hbar\frac{\D}{\D\eta}\langle\tilde{\chi}_{Ks}|
\tilde{\chi}_{Ks}\rangle
\nonumber\\
&=&-\frac{\I S_0'}{m_{\rm P}^2}\int\ud\phi\left[\tilde{\chi}_K\frac{\D}{\D a}
\tilde{\chi}_K+C.C.\right]=0.
\eeqq
To summarize, an appropriate redefinition of the wave functions leads
to a description without unitarity violation. This is why we can
safely neglect the corresponding term in the corrected Schr\"odinger
equation \eqref{eq:3.4}. It is, however, an open issue which of the
wave functions (the original or the redefined one) is the
relevant one in the sense of a quantum measurement process.


\end{document}